# Charge density wave and superconductivity in the family of telluride chalcogenides $Zn_{1-x}Cu_xIr_{2-y}N(N = Al, Ti, Rh)_yTe_4$

Dong Yan[a], Yijie Zeng[b], Yishi Lin[c], Lingyong Zeng[a], Junjie Yin[b], Yuan He[a], Mebrouka Boubeche[a], Meng Wang[b], Yihua Wang[c], Dao-Xin Yao[b], Huixia Luo[a*]

[a]School of Materials Science and Engineering and Key Lab Polymer Composite & Functional Materials, Sun Yat-Sen University, No. 135, Xingang Xi Road, Guangzhou, 510275, P. R. China
[b]School of Physics, State Key Laboratory of Optoelectronic Materials and Technologies, Sun Yat-Sen University, No. 135, Xingang Xi Road, Guangzhou, 510275, P. R. China
[c]Department of Physics, Fudan University, Shanghai, 200433, China

[*]*Corresponding author/authors complete details (Telephone; E-mail:) (+0086)-2039386124*
luohx7@mail.sysu.edu.cn

---

**\*** [*]*Corresponding author/authors complete details (Telephone; E-mail:) (+0086)-2039386124, luohx7@mail.sysu.edu.cn*

**Abstract**

The interplay between superconductivity and charge density wave (CDW)/metal-to-insulator transition (MIT) has long been interested and studied in condensed matter physics. Here we study systematically the charge density wave and superconductivity properties in the solid solutions $Zn_{1-x}Cu_xIr_{2-y}N(N = Al, Ti, Rh)_yTe_4$, which have been successfully synthesized via a solid state method. In contrast with lattice parameters results, we found the lattice parameters, $a$ and $c$, both decrease as Ir-site doping content ($y$) increases, but decrease with the increase of the Cu-site doping content ($x$). Resistivity, magnetic susceptibility and specific heat measurements indicate that the CDW state was suppressed immediately while the superconducting critical temperature ($T_c$) differs from each system. In the Al- and Ti-substitution cases, $T_c$ increase as $y$ increases and reaches a maximum around 2.75 K and 2.84 K respectively at $y = 0.075$, followed by a decrease of $T_c$ before the chemical phase boundary is reached at $y = 0.2$. Nevertheless, $T_c$ decreases monotonously with Rh-doping content $y$ increases and disappears above 0.3 with measuring temperature down to 2 K. Surprisingly, in the $Zn_{1-x}Cu_xIr_2Te_4$ solid solution, $T_c$ enhances as $x$ increases and reaches a maximum value of 2.82 K for $x = 0.5$ but subsequently survives over the whole doping range of $0.00 \leq x \leq 0.9$ despite $T_c$ changes slightly with higher doping content, which differs from the observation of zinc doping suppressing the superconductivity quickly in the high $T_c$ cuprate superconductors. The specific heat anomaly at the superconducting transitions ($\Delta C/\gamma T_{cs}$) for the representative $CuIr_{1.925}Al_{0.075}Te_4$, $CuIr_{1.925}Ti_{0.075}Te_4$ and $Zn_{0.5}Cu_{0.5}Ir_2Te_4$ optimal doping samples are approximately 1.58, 1.44 and 1.45, respectively, which are all slightly higher than the BCS value of 1.43 and indicate bulk superconductivity in these compounds. The results of isothermal magnetization {M(H)} and magneto-transport {$\rho(T, H)$} measurements further suggest that these superconducting compounds are clearly a type-II superconductor. Finally, the CDW transition temperature ($T_{CDW}$) and superconducting transition temperature ($T_c$) vs. $x/y$ content phase diagrams of $Zn_{1-x}Cu_xIr_{2-y}N(N = Al, Ti, Rh)_yTe_4$ have been established and compared, which offers good opportunity to study the competition between CDW and superconductivity in the telluride chalcogenides. Remarkably, in the $Zn_{1-x}Cu_xIr_2Te_4$ system, $T_c$ enhances as $x$ increases and reaches a maximum value of 2.82 K at $x = 0.5$ but subsequently survives over the whole doping range of $0.00 \leq x \leq 0.9$, which is very different from the zinc-doped high $T_c$ cuprate superconductors where zinc doping can kill the superconductivity quickly.

## Introduction

The interplay between superconductivity (SC) and other quantum states such as the spin/charge density wave (S/CDW), metal-to-insulator transition (MIT) or magnetism always attract considerable attention in condensed matter physics due to their exotic and surprising physical properties. In the past, numberless theoretical and experimental scientist have found that the $T_c$ show up when the magnetism can be suppressed or reduced by various dopants in those famous high $T_c$ cuprate or Fe-based superconductors. Although concentrated efforts have been made to gain an understanding of its mechanism between the interplay magnetism and SC, which is still rather unclear. [1-6]

On the other hand, it has been thought that the superconductivity is derived from one kind of Fermi surfaces (FSs) instability due to Cooper pairing, while the CDW state usually appears in low dimensional metallic systems in which the FSs are nested. One class of materials that is of particular interest for study of the interplay between the superconductivity (SC) and charge density wave (CDW) is the two-dimensional transition metal chalcogenides (TMDCs). Previous reports indicate that SC generally competes with CDW in the TMDCs, but the coexistence of SC and CDW states is not uncommon too.[7-8] 2H-NbSe$_2$, for example, has been regarded as one of the most famous layered two-dimensional TMDCs because it has both SC and CDW states with the superconducting transition temperature ($T_c$) of 7.3 K and a quasi-two-dimensional CDW transition temperature ($T_{CDW}$) of ~ 33 K.[9] Besides, the coexistence of SC and CDW states not only emerge in the two-dimensional TMDCs but also in the three-dimensional $MM'_2X_4$ ($M$, $M'$ = transition metal, $X$ = S, Se and Te) type chalcogenides. For instance, CuV$_2$S$_4$ sulfo-spinel is one of the famous three-dimensional spinel materials known to superconductor ($T_c$ = 4.45-3.20 K) and simultaneously exhibits three charge-density wave (CDW) states ($T_{CDW1}$ = 55 K, $T_{CDW2}$ = 75 K, $T_{CDW3}$ = 90 K).[10-11] Metal-to-insulator transition (MIT) is another important electronic quantum state in condensed matter physics. Especially, the CuIr$_2$S$_4$ sulfo-spinel with the same structure with CuV$_2$S$_4$ has been extensively studied due to its the metal-to-insulator transition (MIT) around ~ 230 K, accompanied with a structure transformation. [12-15]

It has been proved that the superconducting, CDW transition and MIT temperatures ($T_{MIT}$) in these aforementioned systems can be tuned by chemical substitution or intercalation.[16-24] 1T-Cu$_x$TiSe$_2$ is a typical example in the two-dimensional TMDC family, in which CDW is suppressed and thereby superconductivity is induced with a maximum $T_c$ of 4.2 K since Cu donates electrons to the pristine 1T-TiSe$_2$ layers. Similarly, the other elements such as Ta, Pd doping TiSe$_2$ also lead to the disappearance of the CDW state and induce superconductivity.[25-26] However, in the other typical CDW-bearing superconducting two-dimensional 2H-NbSe$_2$, it has been observed that the $T_c$ declined as doping content increased no matter intercalation or substitution. Previously, the band calculations for the three-dimensional CuIr$_2$S$_4$ indicate that the electronic states near $E_F$ consist mainly of Ir 5$d$ and S 3$p$ orbitals, while the Cu 3$d$ orbitals form relatively narrow bands.[27] Further, it was experimentally found that the $T_{MIT}$ rapidly decreases with increasing Rh content in

Cu(Ir$_{1-x}$Rh$_x$)$_2$S$_4$,[28] while the MIT was suppressed and superconductivity appears in the Cu$_{1-x}$Zn$_x$Ir$_2$S$_4$ solid solution, with a highest $T_c$ of 3.4 K nearby $x = 0.3$,[29] suggesting that the no matter Cu-site or Ir-site substitution both remarkably alter the electronic states in the CuIr$_2$S$_4$ system.

We have previously reported that CuIr$_2$Te$_4$ was a quasi-two-dimensional ternary telluride chalcogenide superconductor with $T_c \approx 2.5$ K and coexists with a CDW state around 250 K.[30] On the other hand, based on the first principles calculation, we observed the bands of CuIr$_2$Te$_4$ near the Fermi energy $E_F$ mainly come from Te $p$ and Ir $d$ orbitals, similar to that of CuIr$_2$S$_4$ in spinel structure. More recently, we have experimentally proved that both CDW and superconducting properties can be tuned by Ru substitution for Ir in the CuIr$_2$Te$_4$. A "dome-like" shape superconducting transition temperature ($T_c$) vs. x content phase diagram has been established in the CuIr$_{2-x}$Ru$_x$Te$_4$ solid solution, in which a low substitution ($x = 0.03$) of Ru for Ir leads to disappearance of the charge density wave transition, while $T_c$ rises and reaches a maximum value of 2.79 K at $x = 0.05$, followed by a decrease of $T_c$ as x increases. [31] However, how is the interplay between CDW and superconductivity by doping elements in different sites or other dopants with different orbitals is still unknown. Based on these previous reports, we systemically substitute Al, Ti and Rh for Ir and Zn for Cu in the pristine CuIr$_2$Te$_4$ and form the Zn$_{1-x}$Cu$_x$Ir$_{2-y}$N(N = Al, Ti, Rh)$_y$Te$_4$ solid solutions. In this work, we studied the influence of Al ($3p$), Ti ($4s$), Zn ($4s$) and Rh ($5d$) four elements with different orbitals on the superconductivity and CDW state in the pristine CuIr$_2$Te$_4$. We successfully prepared the solid solutions Zn$_x$Cu$_{1-x}$Ir$_2$Te$_4$ ($0.0 \leq x \leq 0.90$) with the Zn substitution for Cu and CuIr$_{2-y}$Al$_y$Te$_4$ ($0.0 \leq y \leq 0.20$), CuIr$_{2-y}$Ti$_y$Te$_4$ ($0.0 \leq y \leq 0.20$), CuIr$_{2-y}$Rh$_y$Te$_4$ ($0.0 \leq y \leq 2.00$) with Al, Ti, Rh substitution for Ir via solid-state reaction method. The structural and electronic properties of Zn$_{1-x}$Cu$_x$Ir$_{2-y}$N(N = Al, Ti, Rh)$_y$Te$_4$ were characterized via X-ray diffraction (XRD), temperature-dependent resistivity, magnetic susceptibility and specific heat measurements. As a result, the structural and physical properties are strongly influenced by these chemical doping. In all cases, the CDW state has been immediately suppressed via using Al, Ti, Rh and Zn doping as finely controlled tuning parameters. However, the superconducting critical temperature ($T_c$) changes in different trends. In the Al- and Ti- substitution systems, $T_c$ increase as y increases and reaches a maximum around 2.75 K and 2.84 K respectively for $y = 0.075$. Nevertheless, $T_c$ decreases monotonously with Rh-doping content y increases and disappears at $y = 0.3$ above 2 K. Surprisingly, in the Zn$_{1-x}$Cu$_x$Ir$_2$Te$_4$ solid solution, $T_c$ enhances as x increases and reaches a maximum value of 2.82 K for $x = 0.5$ but is observed over the whole doping range of $0.00 \leq x \leq 0.9$ despite $T_c$ changes slightly with higher doping content. However, we find that the superconducting transition temperature ($T_c$) of the similarly made and tested isostructural CuIr$_{2-y}$Rh$_y$Te$_4$ decreases monotonously with y increases and disappears at $x = 0.3$ above 2 K. Finally, we present a comparison of the electronic phase diagrams of many doped CuIr$_2$Te$_4$ systems, showing that they behave quite differently, which may have broad implications in the search for new superconductors or be suitable material platform candidates for further study of the interplay between CDW and superconductivity or compare the interplay between magnetism and SC. From these facts, we conclude that the superconducting mechanism in the $Zn_{1-x}Cu_xIr_2Te_4$ is very different from the zinc-doped high $T_c$ cuprate superconductors where zinc doping can kill the superconductivity quickly.[1-4] In the cuprate superconductors, the SDW or spin fluctuations is thought to play key role in the superconducting mechanism since the zinc ions do not have spin when they substitute the S=1/2 $Cu^{2+}$ ions.

**Experimental Section**

Polycrystalline samples with the formula $CuIr_{2-y}Al_yTe_4$ (0.0 ≤ y ≤ 0.20), $CuIr_{2-y}Ti_yTe_4$ (0.0 ≤ y ≤ 0.20), $CuIr_{2-y}Rh_yTe_4$ (0.0 ≤ y ≤ 2.00) and $Zn_xCu_{1-x}Ir_2Te_4$ (0.0 ≤ x ≤ 0.90) were synthesized in two steps by a solid-state reaction method. First, the mixture of high purity fine powders of Cu (99.9 %), Ir (99.9 %), Al (99.999 %), Ti (99.999 %), Zn (99.999 %) and Te (99.999 %) in the appropriate stoichiometric ratios were heated in sealed evacuated silica glass tubes at a rate of 1 °C/min to 850 °C and held there for 120 hours. Subsequently, the as-prepared powders were reground, re-pelletized and sintered again, by heating at a rate of 3 °C/min to 850 °C and holding there for 96 hours. For some samples, several ground, pelletized and sintered are need. The identity and phase purity of the samples were determined by powder X-ray diffraction (PXRD) using Rigaku with Cu Kα radiation and a LYNXEYE-XE detector. To determine the unit cell parameters, profile fits were performed on the powder diffraction data in the FULLPROF diffraction suite using Thompson-Cox-Hastings pseudo-Voigt peak shapes modle.[32] Measurements of the temperature dependent electrical resistivity (4-point method), specific heat, and magnetic susceptibility of the materials were performed in a DynaCool Quantum Design Physical Property Measurement System (PPMS). There was no indication of air-sensitivity of the materials during the study. $T_c$ determined from susceptibility data were estimated conservatively: $T_c$ was taken as the intersection of the extrapolations of the steepest slope of the susceptibility in the superconducting transition region and the normal state susceptibility; for resistivities, the midpoint of the resistivity $ρ(T)$ transitions was taken, and for the specific heat data, the critical temperatures obtained from the equal area construction method were employed.

**Results and Discussion**

**Fig. 1** displays the X-ray diffraction (XRD) patterns for the polycrystalline $Zn_{1-x}Cu_xIr_{2-y}N(N = Al, Ti, Rh)_yTe_4$ samples at room temperature. XRD results indicate that the solubility limit for Al, Ti, Rh and Zn doping in $CuIr_2Te_4$ is 0.2, 0.2, 2.0 and 0.90, respectively. With higher doping contents, the $Al_2Te_3$, $TiTe_2$ and ZnTe phases are obviously found as impurities, respectively. Besides, the enlargement of (001) peak in **Fig. 1** shows obvious right shift with the increasing contents of Al, Ti, and Rh, while the (001) peak exhibits left shift with the increasing contents of Zn. This phenomenon was also according with the evolution of fitting unit cell parameters $c$ in **Fig. S1** by the means of crystal plane spacing formula.

**Fig. S1** shows the powder X-ray diffraction patterns at room temperature and fitting unit cell parameters for the representative $CuIr_{1.925}Al_{0.075}Te_4$, $CuIr_{1.925}Ti_{0.075}Te_4$, $CuIr_{1.95}Rh_{0.05}Te_4$ and $Zn_{0.5}Cu_{0.5}Ir_2Te_4$ samples. **Fig. S1** main panel shows the detail refinement results of the selected powders. Most of the reflections can be indexed in the $P\bar{3}m1$ space group and the tiny impurity is attributed to the unreacted Ir. The inset pattern shows that each system adopts a disordered trigonal structure with a space group $P\bar{3}m1$, which embodies a two-dimensional (2D) $IrTe_2$ layers and intercalated by Cu or Cu/Zn between the layers, with Ir partial replacing by Al, Ti and Rh, respectively. The lattice constants $a$ and $c$ for the solid solution $Zn_{1-x}Cu_xIr_{2-y}N(N = Al, Ti, Rh)_yTe_4$ vary systematically with increasing doping concentration $x$ or $y$, as shown in **Fig. 2**. Obviously, all the lattice constants $a$ and $c$ for our studied systems change linearly as doping contents increase, which obey the Vegard's law. Additionally, the unit cell parameters $a$ and $c$ both decrease linearly with the increase of Ti-, Al- and Rh-concentration $y$. On the contrary, $a$ and $c$ parameters are both increase as Zn contents increase. For example, lattice constant $a$ decrease from 3.9397(5) Å ($y = 0$) to 3.9264(1) Å ($y = 0.20$) and $c$ decreases from 5.3965 (3) Å ($y = 0$) to 5.3757 (2) Å ($y = 0.20$) in $CuIr_{2-y}Al_yTe_4$. In the case of $CuIr_{2-y}Ti_yTe_4$, unit cell parameter $a$ decline from 3.9397(5) Å ($y = 0$) to 3.9310(2) Å ($y = 0.20$) and $c$ decreased from 5.3965 (3) Å ($y = 0$) to 5.3792 (2) Å ($y = 0.20$). And in the solid solution $CuIr_{2-y}Rh_yTe_4$, $a$ decline from 3.9397(5) Å ($y = 0$) to 3.7668(2) Å ($y = 2.00$) and $c$ decreased from 5.3965 (3) Å ($y = 0$) to 5.2704 (5) Å ($y = 2.00$). Yet, unit cell parameter $a$ rise from 3.9397(5) Å ($y = 0$) to 3.9553(3) Å ($y = 0.90$) and $c$ increased from 5.3965 (3) Å ($y = 0$) to 5.4435 (6) Å ($y = 0.90$) for $Zn_xCu_{1-x}Ir_2Te_4$.

Hereafter, we focus on the temperature dependence of the electrical resistivity $\rho(T)$ and magnetic susceptibility $M(T)$ measurements to examine the conducting and superconducting properties of investigated compounds $Zn_{1-x}Cu_xIr_{2-y}N(N = Al, Ti, Rh)_yTe_4$. **Fig. 3** displays the temperature dependence of the normalized electrical resistivities ($\rho/\rho_{300K}$) for the polycrystalline samples of $CuIr_{2-y}Al_yTe_4$ ($0.0 \leq y \leq 0.20$), $CuIr_{2-y}Ti_yTe_4$ ($0.0 \leq y \leq 0.20$), $CuIr_{2-y}Rh_yTe_4$ ($0.0 \leq y \leq 2.00$) and $Zn_xCu_{1-x}Ir_2Te_4$ ($0.0 \leq x \leq 0.90$) in the range of 2 - 300 K. As shown in **Fig. 4**, an obvious and steep drops of $\rho(T)$ can be seen in the $CuIr_{2-y}Al_yTe_4$ ($0.0 \leq y \leq 0.15$), $CuIr_{2-x}Ti_yTe_4$ ($0.0 \leq y \leq 0.15$), $CuIr_{2-y}Rh_yTe_4$ ($0.0 \leq y \leq 0.20$) and $Zn_xCu_{1-x}Ir_2Te_4$ ($0.0 \leq x \leq 0.90$) at low temperatures, signifying the onset of superconductivity at low temperatures (1.6 - 3.2 K). However, with higher Al-, Ti- and Rh-concentration $y$, although the $\rho(T)$ decreases with the decrease of measuring temperature, the obvious and steep drops of $\rho(T)$ begin to vanish, indicating the $CuIr_{2-y}Al_yTe_4$ ($0.15 < y \leq 0.2$), $CuIr_{2-y}Ti_yTe_4$ ($0.15 < y \leq 0.2$), $CuIr_{2-y}Rh_yTe_4$ ($0.2 < y \leq 0.3$) became metallic phase above 2 K. Besides, it can found that the hump of $\rho(T)$ disappears even with slightly substitution concentration in all $Zn_{1-x}Cu_xIr_{2-y}N(N = Al, Ti, Rh)_yTe_4$ samples, suggesting the CDW has been suppressed by doping, which is similar to the behavior in our previous study of $CuIr_{2-x}Ru_xTe_4$.[31] The superconducting transitions were further confirmed by the magnetic susceptibility measurements. Based on the $\rho(T)$ results, we further performed the magnetic susceptibility measurements for the selected $CuIr_{2-y}Al_yTe_4$ ($0.0 \leq y \leq 0.2$), $CuIr_{2-x}Ti_yTe_4$ ($0.0 \leq y \leq 0.2$), $CuIr_{2-y}Rh_yTe_4$ ($0.0 \leq y \leq 0.20$) and $Zn_xCu_{1-x}Ir_2Te_4$ ($0.0 \leq x \leq 0.90$) samples. As shown in **Fig. 5,** the onset of the negative magnetic susceptibility signals the systematical superconducting state for $CuIr_{2-y}Al_yTe_4$ ($0.0 \leq y \leq 0.15$), $CuIr_{2-y}Ti_yTe_4$ ($0.0 \leq y \leq 0.15$), $CuIr_{2-y}Rh_yTe_4$ ($0.0 \leq y \leq 0.20$) and $Zn_xCu_{1-x}Ir_2Te_4$ ($0.0 \leq x \leq 0.90$), respectively. We also found that the negative magnetic susceptibility signals disappear for Ti- and Al- concentration $y = 0.2$ samples, further indicating that the $T_c$ was suppressed with higher Ti- or Al-doping content, which is consistent with the $\rho(T)$ results. In additional, we can estimate the superconducting volume fraction for all superconducting samples approximately to be 96 %, revealing the high purity of the polycrystalline $Zn_{1-x}Cu_xIr_{2-y}N(N = Al, Ti, Rh)_yTe_4$ samples.

In order to determine the lower critical field $\mu_0H_{c1}(0)$, we next performed the temperature-dependent measurements of the magnetization under incremental magnetic field M(H). We choose the optimal doping-concentration compounds of each system for further studied. **Fig. 6** shows how to calculate the lower critical filed $\mu_0H_{c1}(0)$ for the representative optimal $CuIr_{1.925}Al_{0.075}Te_4$, $CuIr_{1.925}Ti_{0.075}Te_4$, $CuIr_{1.95}Rh_{0.05}Te_4$ and $Zn_{0.5}Cu_{0.5}Ir_2Te_4$, respectively. Now we describe in detail how to measure and get the $\mu_0H_{c1}(0)$ for the $CuIr_{1.925}Ti_{0.075}Te_4$ compound as an example. First, we performed applied field magnetization M(H) measurements at different low temperatures (1.8 K, 2.0 K, 2.2 K, 2.4 K and 2.6 K) to further calculate the demagnetization factor (N). Assuming that the beginning linear response to the magnetic field is perfectly diamagnetic (dM/dH = −1/4π) for this bulk superconductor, the values of demagnetization factor (N) was determined to be 0.41 – 0.53 (from N = 1/4π $\chi_V$ + 1), where $\chi_V$ = dM/dH is the value of linearly fitted slope for the up right corner inset of **Fig. 6b**. We further can fit the experimental data according to the equation $M_{fit}$ = $a$ + $b$H at low magnetic fields, where $a$ stands for an intercept and $b$ notes as a slope from fitting the low magnetic field magnetization measurements data. The plot for the M(H) − $M_{fit}$ data versus the magnetic field(H) is shown in the bottom-left corner inset of **Fig. 6**. Further, the $\mu_0H_{c1}^*$ was obtained at the field when M deviates by ∼ 1 % above the fitted data ($M_{fit}$), as is the common practice.[31,33] Then we can determine the lower critical field $\mu_0H_{c1}(T)$ *via* using the relation $\mu_0H_{c1}(T)$ = $\mu_0H_{c1}^*(T)$/ (1 − N), with considering of the demagnetization factor (N). [34,35] The obtained $\mu_0H_{c1}(T)$ as the function of temperature for $CuIr_{1.925}Ti_{0.075}Te_4$ was plotted in the main panel of **Fig. 6b**. The $\mu_0H_{c1}(0)$ can be further determined by fitting the $\mu_0H_{c1}(T)$ data based on the equation $\mu_0H_{c1}(T) = \mu_0H_{c1}(0) [1 − (T/T_c)^2]$, which was shown by the black solid lines. The obtained zero-temperature lower critical field $\mu_0H_{c1}(0)$ for $CuIr_{1.925}Al_{0.075}Te_4$, $CuIr_{1.925}Ti_{0.075}Te_4$, $CuIr_{1.95}Rh_{0.05}Te_4$ and $Zn_{0.5}Cu_{0.5}Ir_2Te_4$ was 0.060, 0.095, 0.045 and 0.062 T, respectively (**Table 1**), which is higher than that of the host compound $CuIr_2Te_4$.

We now discuss the evolution of temperature dependent electrical resistivity at various applied fields $\rho(T, H)$ for the optimal doping-concentration compounds $CuIr_{1.925}Al_{0.075}Te_4$, $CuIr_{1.925}Ti_{0.075}Te_4$, $CuIr_{1.95}Rh_{0.05}Te_4$ and $Zn_{0.5}Cu_{0.5}Ir_2Te_4$ to estimate the upper critical field $\mu_0H_{c2}(0)$, as shown in **Fig. 7**. The upper critical field values $\mu_0H_{c2}$ vs superconducting temperature transition $T_c$ determined from resistivity at different applied fields was plotted in the insets of **Fig. 7**. It can be seen that all the $\mu_0H_{c2}$ vs T curves near $T_c$ of $CuIr_{1.925}Al_{0.075}Te_4$, $CuIr_{1.925}Ti_{0.075}Te_4$, $CuIr_{1.95}Rh_{0.05}Te_4$ and $Zn_{0.5}Cu_{0.5}Ir_2Te_4$ samples exhibit the well linearly fitting, which are represented by solid line. The values of fitting data slope (d$H_{c2}$/d$T$) of $CuIr_{1.925}Al_{0.075}Te_4$, $CuIr_{1.925}Ti_{0.075}Te_4$,

$CuIr_{1.95}Rh_{0.05}Te_4$ and $Zn_{0.5}Cu_{0.5}Ir_2Te_4$ samples were listed in **Table 1**. Inspection of $\rho(T, H)$, the zero-temperature upper critical field ($\mu_0H_{c2}(T)$) for $CuIr_{1.925}Al_{0.075}Te_4$, $CuIr_{1.925}Ti_{0.075}Te_4$, $CuIr_{1.95}Rh_{0.05}Te_4$ and $Zn_{0.5}Cu_{0.5}Ir_2Te_4$ was calculated to be 0.202, 0.212, 0.167 and 0.198 T, respectively, according to the Werthamer-Helfand-Hohenberg (WHH) expression equation $\mu_0H_{c2}(T) = -0.693T_c (dH_{c2}/dT_c)$ for the dirty limit superconductivity.[31, 34-35] The obtained $\mu_0H_{c2}(T)$ for each superconductor, as compared with that of the pristine $CuIr_2Te_4$, is much higher. In addition, the Pauli limiting field ($\mu_0H^P(T)$) of $CuIr_{1.925}Al_{0.075}Te_4$, $CuIr_{1.925}Ti_{0.075}Te_4$, $CuIr_{1.95}Rh_{0.05}Te_4$ and $Zn_{0.5}Cu_{0.5}Ir_2Te_4$ can be calculated from $\mu_0H^P(T) = 1.86T_c$. Then, with this formula $\mu_0H_{c2}(T) = \frac{\phi_0}{2\pi\xi_{GL}^2}$, where $\phi_0$ is the flux quantum, the Ginzburg-Laudau coherence length ($\xi_{GL}(0)$) was calculated to be around 40.4, 39.3, 44.4, and 40.7 nm for $CuIr_{1.925}Al_{0.075}Te_4$, $CuIr_{1.925}Ti_{0.075}Te_4$, $CuIr_{1.95}Rh_{0.05}Te_4$ and $Zn_{0.5}Cu_{0.5}Ir_2Te_4$, respectively. All the relative superconducting parameters were summarized in **Table 1**.

To further confirm that superconductivity is an intrinsic property of $CuIr_{1.925}Al_{0.075}Te_4$, $CuIr_{1.925}Ti_{0.075}Te_4$ and $Zn_{0.5}Cu_{0.5}Ir_2Te_4$, the temperature-dependent specific heat measurements were carried out with the exception of magnetic susceptibility and resistivity measurements. Herein, we present the detailed process for $CuIr_{1.925}Ti_{0.075}Te_4$ as an example. In order to estimate $\beta$ and $\gamma$ parameters, we potted $C_p/T$ vs. $T^2$ in **the inset of Fig. 8b.** By fitting above the critical temperature to the $C_p = \gamma T + \beta T^3$, where $\beta T^3$ represents the phonon contribution ($C_{ph.}$) and $\gamma T$ is the normal-state electronic contribution ($C_{el.}$) to the specific heat, the value of $\beta$ is approximated to be 2.71 mJ mol$^{-1}$ K$^{-4}$ and the extrapolation to $T = 0$ gives $\gamma = 13.94$ mJ mol$^{-1}$ K$^{-2}$ for $CuIr_{1.925}Ti_{0.075}Te_4$. **The main panel of Fig. 8b** displays the electronic specific heat divided by temperature ($C_{el.}/T$) vs. T for $CuIr_{1.925}Al_{0.075}Te_4$ in the range of 2.0 - 4.5 K under zero magnetic field, where $C_{el.}$ was obtained by subtracting the phonon contribution to the specific heat: $C_{el.} = C_p - \beta T^3$. It was obviously seen that a large jump occurred in the specific heat. Further, we used an equal-area entropy construction (red solid lines) method to determine the $T_c$ to be 2.80 K for $CuIr_{1.925}Ti_{0.075}Te_4$, which is consistent with the $T_c$s from our resistivity and magnetic susceptibility measurements. Further, we obtained the normalized specific heat jump value $\Delta C/\gamma T_c$ from entropy conservation construction in the **inset of Fig. 8a** to be 1.44 for $CuIr_{1.925}Ti_{0.075}Te_4$, which is very close to the that of the Bardeen-Cooper-Schrieffer (BCS) weak-coupling limit value (1.43), confirming bulk superconductivity. In addition, the Debye temperatures can be calculated by the formula $\Theta_D = (12\pi^4 nR/5\beta)^{1/3}$ by using the fitted value of $\beta$, where $n$ is the number of atoms per formula unit and $R$ is the gas constant. Subsequently, the electron-phonon coupling constant ($\lambda_{ep}$) is calculated to be 0.68 by using the Debye temperature ($\Theta_D$) and critical temperature $T_c$ from the inverted McMillan formula: $\lambda_{ep} = \frac{1.04 + \mu^* \ln\left(\frac{\Theta_D}{1.45T_c}\right)}{(1-1.62\mu^*)\ln\left(\frac{\Theta_D}{1.45T_c}\right) - 1.04}$ [36]. Further, we can estimate the electron density of states at the Fermi level ($N(E_F)$) by using the equation $N(E_F) = \frac{3}{\pi^2 k_B^2 (1+\lambda_{ep})}\gamma$ with the $\gamma$ and $\lambda_{ep}$. The determination gives $N(E_F) = 3.52$ states/eV f.u. for $CuIr_{1.925}Ti_{0.075}Te_4$, which is higher than that ($N(E_F)$ = 3.10 states/eV f.u.) of $CuIr_2Te_4$ (**Table 1**). Using the aforementioned process, we can get the relative superconducting parameters for the other two optimal $CuIr_{1.925}Al_{0.075}Te_4$ and $Zn_{0.5}Cu_{0.5}Ir_2Te_4$ samples, which have been summarized in **Table 1**.

Lastly, to further understand the influence of different doptants on the CDW and superconductivity, the overall behavior of this system is summarized in the electronic phase diagram presented in **Fig. 9**. **Fig. 9** presents the electronic phase diagrams plotted $T_c$ verus doping-concentration $x/y$ for $Zn_{1-x}Cu_xIr_{2-y}N(N = Al, Ti, Rh)_yTe_4$. In this figure, all the $T_c$ were determined from the temperature dependence of the normalized ($\rho/\rho_{300K}$) resistivities. Using Al, Ti, Rh or Zn doping as a finely controlled tuning parameter, the charge density wave state was suppressed immediately while the superconducting critical temperature ($T_c$) was altered in different trends. In the cases of Al- and Ti-doping, $T_c$ increase as $y$ increases and reaches a maximum $T_c$ of 2.75 K and 2.84 K, respectively at $y$ = 0.075, followed by a decline of $T_c$ before the chemical solubility limit is reached at $y$ = 0.2. Nevertheless, $T_c$ decreases monotonously with Rh-doping content $y$ increases and disappears around $y$ = 0.3 with measuring temperature down to 2 K. It should be noted here that by adjusting the Zn-concentration $x$ in the $Zn_{1-x}Cu_xIr_2Te_4$ solid solution, $T_c$ enhances as $x$ increases and goes through a maximum value of 2.82 K at $x$ = 0.5, but subsequently exists in the whole doping range of $0.00 \leq x \leq 0.9$ despite $T_c$ changes slightly with higher Zn-concentration $x$. The comparison of the phase diagrams signifies that even if a small amount of Al, Ti and Rh atoms is substituted for Ir, the superconducting properties are strongly influenced by this gentle chemical pressure. However, the $T_c$ changes slightly as Zn increases in $Zn_xCu_{1-x}Ir_2Te_4$. What is reason behind this behavior? On the basis of our calculation on $CuIr_2Te_4$, we have previously found both orbital projected band structure and density of state, the bands near the Fermi energy $E_F$ mainly come from Te $p$ and Ir $d$ orbitals. Therefore, we propose that chemical substituted for Ir-site in the $CuIr_2Te_4$ maybe play a more important role on the chemical substitution for Cu-site in the host $CuIr_2Te_4$. Yet, more studied need to be done to prove it. On the other hand, the reason why the CDW state can be suppressed by chemical doping so quickly has not yet been studied. Through the systematic study of the $Zn_{1-x}Cu_xIr_{2-y}N(N = Al, Ti, Rh)_yTe_4$, we hope that these experimental results will be useful for discovery of new superconductors and the clarification of the interplay between CDW and superconductivity.

**Conclusion**

Here the solid solutions $Zn_{1-x}Cu_xIr_{2-y}N(N = Al, Ti, Rh)_yTe_4$ have been successfully synthesized via the solid-state reaction. The structural and superconductivity properties for this system were evaluated systematically by means of powder X-ray diffraction (PXRD), magnetization, resistivity and specific heat measurements. Our results indicate that the lattice parameters, $a$ and $c$, both decrease with increasing Ir-site doping content, but decrease with increasing Cu-site doping content, which obey the Vegard's law. Resistivity, magnetic susceptibility and specific heat measurements indicate that the CDW state was suppressed immediately while the superconducting critical temperature ($T_c$) changed very different in these four systems. Superconducting temperature ($T_c$) increases as $y$ increases and reaches a maximum $T_c$ of 2.75 K and 2.84 K respectively with doping content 0.075, in both Al- and Ti-substitution cases. However, $T_c$ declines continuously with the increase of Rh-doping content and disappears above $y = 0.3$ with measuring temperature down to 2 K. Remarkably, in the $Zn_{1-x}Cu_xIr_2Te_4$ system, $T_c$ enhances as $x$ increases and reaches a maximum value of 2.82 K at $x = 0.5$ but subsequently survives over the whole doping range of $0.00 \leq x \leq 0.9$ despite $T_c$ changes slightly with higher doping content. This is very different from the zinc-doped high $T_c$ cuprate superconductors where zinc doping can kill the superconductivity quickly. The specific heat anomaly at the superconducting transitions ($\Delta C/\gamma T_{cs}$) for the representative $CuIr_{1.925}Al_{0.075}Te_4$, $CuIr_{1.925}Ti_{0.075}Te_4$ and $Zn_{0.5}Cu_{0.5}Ir_2Te_4$ otipmal doping samples are approximately 1.58, 1.44 and 1.45, respectively, which are all slightly higher than the BCS value of 1.43 and indicate bulk superconductivity in these compounds. The results of isothermal magnetization {M(H)} and magneto-transport {$\rho(T, H)$} measurements further suggest that these superconducting compounds are clearly a type-II superconductor. Finally, the CDW transition temperature ($T_{CDW}$) and superconducting transition temperature ($T_c$) vs. $x/y$ content phase diagrams of these doped systems have been established and compared, which gives an opportunity for the further study of the competition between CDW and the superconductivity in the telluride chalcogenides. It also maybe offers a good platform to compare the SDW/CDW and superconductivity in the high $T_c$ systems.

**ACKNOWLEDGMENT**
The authors thank B. Shen for valuable discussions. H. X. Luo acknowledges the financial support by the National Natural Science Foundation of China (Grants No.11922415 and No.21701197) and the Fundamental Research Funds for the Central Universities (19lgzd03), Guangdong Basic and Applied Basic Research Foundation (2019A1515011718), and the Pearl River Scholarship Program of Guangdong Province Universities and Colleges (20191001). Y. Zeng and D. X. Yao are supported by NKRDPC Grants No. 2017YFA0206203, No. 2018YFA0306001, NSFC-11574404, National Supercomputer Center in Guangzhou, and Leading Talent Program of Guangdong Special Projects. D. Yan acknowledges the financial support by National Key Laboratory Development Fund (No. 20190030). M. Wang and J. J. Yin are supported by NSFC-11904414, and NSF of Guangdong (Grant No. 2018A030313055).

## References

1. Qi, X. L.; Weng, Z. Y. Microscopic origin of local moments in a zinc-doped high-$T_c$ superconductor. *Phys. Rev. B* **2005**, *71*, 184507.
2. Finkel'stein, A. M.; Kataev, V. E.; Kukovitskii, E. F.; Teitel'Baum, G. B. Effects of Zn substitution for Cu atoms in lanthanum-strontium superconductors. *Phys. C*

**1990**, *168*, 370-380.

3. Lanckbeen, A.; Duvigneaud, P. H.; Diko, P.; Mehbod, M.; Naessen, G.; Deltour, R. Effect of zinc and iron on the (micro-) structure and copper charge excess of the YBaCuO superconductor. *J. Mater. Sci.* **1994**, *29*, 5441-5448.

4. Pan, S. H.; Hudson, E. W.; Lang, K. M.; Eisaki, H.; Uchida, S.; Davis, J. C. Imaging the effects of individual zinc impurity atoms on superconductivity in $Bi_2Sr_2CaCu_2O_{8+\delta}$. *Nature*, **2000**, *403*, 746-750.

5. Li, J.; Guo, Y.; Zhang, S.; Yu, S.; Tsujimoto, Y.; Kontani, H.; Takayama-Muromachi, E. Linear decrease of critical temperature with increasing Zn substitution in the iron-based superconductor $BaFe_{1.89-2x}Zn_{2x}Co_{0.11}As_2$. *Phys. Rev. B* **2011**, *84*, 020513.

6. Yao, Z. J.; Chen, W. Q.; Li, Y. K.; Cao, G. H.; Jiang, H. M.; Wang, Q. E.; Zhang, F. C. Zn-impurity effect and interplay of $s_\pm$ and $s_{++}$ pairings in iron-based superconductors. *Phys. Rev. B* **2012**, *86*, 184515.

7. Wilson, J. A.; Di Salvo, F. J.; Mahajan, S. Charge-density waves and superlattices in the metallic layered transition metal dichalcogenides. *Adv. Phys.* **1975**, *24*, 117-201.

8. Gabovich, A. M.; Voitenko, A. I.; Ausloos, M. Charge-and spin-density waves in existing superconductors: competition between Cooper pairing and Peierls or excitonic instabilities. *Phys. Rep.* **2002**, *367*, 583-709.

9. Revolinsky, E.; Lautenschlager, E. P.; Armitage, C. H. Layer structure superconductor. *Solid State Common.* **1963**, *1*, 59-61.

10. Fleming, R. M.; DiSalvo, F. J.; Cava, R. J.; Waszczak, J. V. Observation of charge-density waves in the cubic spinel structure $CuV_2S_4$. *Phys. Rev. B* **1981**, *24*, 2850-2853.

11. Nagard, N. L.; Katty, A.; Collin, G.; Gorochov, O.; Willig, A. Preparation, crystal structure and physical properties (transport, magnetic susceptibility and NMR) of $CuV_2S_4$ spinel. *J. Solid State Chem.* **1979**, *27*, 267-277.

12. Nagata, S.; Hagino, T.; Seki, Y.; Bitoh, T. Metal-insulator transition in thiospinel $CuIr_2S_4$. *Phys. B* **1994**, *1077*, 194–196.

13. Furubayashi, T.; Matsumoto, T.; Hagino, T.; Nagata, S. Structural and magnetic studies of metal-insulator transition in thiospinel $CuIr_2S_4$. *J. Phys. Soc. Jpn.* **1994**, *63*, 3333-3339.

14. Radaelli, P. G.; Horibe, Y.; Gutmann, M. J.; Ishibashi, H.; Chen, C. H.; Ibberson, R. M.; Koyama, Y.; Hor, Y. S.; Kiryukhin, V.; Cheong, S. W. Formation of isomorphic $Ir^{3+}$ and $Ir^{4+}$ octamers and spin dimerization in the spinel $CuIr_2S_4$. *Nature* **2002**, *416*, 155-158.

15. Oda, T.; Shirai, M.; Suzuki, N.; Motizuki, K. Electronic band structure of sulphide spinels $CuM_2S_4$ (M= Co, Rh, Ir). *J. Phys.: Condens. Matter* **1995**, *7*, 4433.

16. Luo, H. X.; Strychalska-Nowak, J.; Li, J.; Tao, J.; Klimczuk, T.; Cava, R. J. S-Shaped Suppression of the Superconducting Transition Temperature in Cu-Intercalated $NbSe_2$. *Chem. Mater.* **2017**, *29*, 3704-3712.

17. Hauser, J. J.; Robbins, M.; DiSalvo, F. J. Effect of 3d impurities on the superconducting transition temperature of the layered compound $NbSe_2$. *Phys. Rev. B* **1973**, *8*, 1038-1042.

18. Voorhoeve-Van Den Berg, J. M. On the ternary phases Al$_x$NbSe$_2$ and Cu$_x$NbSe$_2$. *J. Less-Common. Metals*. **1972**, *26*, 399-402.
19. Anderson, P. W. The theroy of dirty of superconductors. *J. Phys. Chem. Solids* **1959**, *11*, 26-30.
20. Sugawara, K.; Yokota, K.; Tanokura, Y.; Sekine, T. Anisotropy of flux flow in layered superconductor 2H-NbSe$_{2-x}$S$_x$. *J. Low Temp. Phys.* **1993**, *90*, 305-317.
21. Sugawara, K.; Yokota, K.; Takei, N.; Tanokura, Y.; Sekine, T. Dimension of the collective pinning in the layered superconductor 2H-NbSe$_{2-x}$S$_x$. *Phys. B* **1994**, *194-196*, 1891-1892.
22. Naito, M.; Tanaka, S. Electrical Transport Properties in 2H-NbS$_2$, -NbSe$_2$, -TaS$_2$ and -TaSe$_2$. *J. Phys. Soc. Jpn.* **1982**, *51*, 219-227.
23. Yan, D.; Lin, Y. S.; Wang, G. H.; Zhu, Z.; Wang, S.; Shi, L.; He, Y.; Li, M. R.; Zheng, H.; Ma, J.; Jia, J. F.; Wang, Y. H.; Luo, H. X. The Unusual Suppression of Superconducting Transition Temperature in Double-Doping 2H-NbSe$_2$, *Supercond. Sci. Tech*. **2019**, *32*, 085008.
24. Yan, D.; Wang, S.; Lin, Y. S.; Wang, G. H.; Zeng, Y. J.; Boubeche, M.; He, Y.; Ma, J.; Wang, Y. H.; Yao, D. X.; Luo, H. X. NbSeTe — A new layered transition metal dichalcogenide superconductor, *J. Phys.: Condens. Matter* **2020**, *32*, 025702
25. Morosan, E.; Zandbergen, H. W.; Dennis, B. S.; Bos, J. W. G.; Onose, Y.; Klimczuk, T.; Ramirez, A. P.; Ong, N. P.; Cava, R. J. Superconductivity in Cu$_x$TiSe$_2$. *Nat. Phys.* **2006**, *2*, 544-550.
26. Barath, H.; Kim, M.; Karpus, J. F.; Cooper, S. L.; Abbamonte, P.; Fradkin, E.; Morosan, E.; Cava, R. J. Quantum and classical mode softening near the charge-density-wave-superconductor transition of Cu$_x$TiSe$_2$. *Phys. Rev. Lett.* **2008**, *100*, 106402.
27. Oda, T.; Shirai, M.; Suzuki, N.; Motizuki, K. Electronic band structure of sulphide spinels CuM$_2$S$_4$ (M= Co, Rh, Ir). *J. Phys.: Condens. Matter* **1995**, *7*, 4433.
28. Matsumoto, N.; Endoh, R.; Nagata, S.; Furubayashi, T.; Matsumoto, T. Metal-insulator transition and superconductivity in the spinel-type Cu(Ir$_{1-x}$Rh$_x$)$_2$S$_4$ system. *Phys. Rev. B* **1999**, *60*, 5258.
29. Suzuki, H.; Furubayashi, T.; Cao, G.; Kitazawa, H.; Kamimura, A.; Hirata, K.; Matsumoto, T. Metal-Insulator Transition and Superconductivity in Spinel-Type System Cu$_{1-x}$Zn$_x$Ir$_2$S$_4$. *J Phys. Soc. Jpn.* **1999**, *68*, 2495-2497.
30. Yan, D.; Zeng, Y. J.; Wang, G. H.; Liu, Y. Y.; Yin, J. J.; Chang, T.-R.; Lin, H.; Wang, M.; Ma, J.; Jia, S.; Yao, D.-X.; Luo, H. X. *arXiv:1908.05438*.
31. Yan, D.; Zeng, L.; Lin, Y.; Yin, J.; He, Y.; Zhang, X.; Huang, M.; Shen, B.; Wang, M.; Wang, Y.; Yao, D.-X.; Luo, H. X. Superconductivity in Ru-doped CuIr$_2$Te$_4$ telluride chalcogenide, *Phys. Rev. B* **2019,** *100*, 174504.
32. Rodríguez-Carvajal, J. FULLPROF: A Program for Rietveld Refinement and Pattern Matching Analysis. *Comm. Powder Diffr.* **2001**, *26*, 12-70.
33. Luo, H. X.; Xie, W. W.; Tao, J.; Pletikosic, I.; Valla, T. G.; Sahasrabudhe, S.; Osterhoudt, G.; Sutton, E.; Burch, K. S.; Seibel, E. M.; Krizan, J. W.; Zhu, Y. M.; Cava, R. J. Differences in Chemical Doping Matter: Superconductivity in Ti$_{1-x}$Ta$_x$Se$_2$ but Not in Ti$_{1-x}$Nb$_x$Se$_2$. Chem. Mater. 2016, 28, 1927-1935.

34. Winiarski, M. J.; Wiendlocha, B.; Gołąb, S.; Kushwaha, S. K.; Wiśniewsk, P.; Kaczorowski, D.; Thompson, J. D.; Cava, R. J.; Klimczuk, T. Superconductivity in CaBi$_2$. *Phys. Chem. Chem. Phys.* **2016**, *18*, 21737.
35. Yadav, C. S.; Paulose, P. L. Upper critical field, lower critical field and critical current density of FeTe$_{0.60}$Se$_{0.40}$ single crystals. *New J. Phys.* **2009**, *11*, 103046.
36. McMillan, W. L. Transition Temperature of Strong-Coupled Superconductors. *Phys. Rev.* **1968**, *167*, 331-344.

**Table 1. Comparison of superconducting parameters in ternary/quaternary telluride chalcogenides.**

| Material | CuIr$_{1.925}$Al$_{0.075}$Te$_4$ | CuIr$_{1.925}$Ti$_{0.075}$Te$_4$ | CuIr$_{1.95}$Rh$_{0.05}$Te$_4$ | Zn$_{0.5}$Cu$_{0.5}$Ir$_2$Te$_4$ | CuIr$_{1.95}$Ru$_{0.05}$Te$_4$ | CuIr$_2$Te$_4$ |
|---|---|---|---|---|---|---|
| $T_c$ (K) | 2.75 | 2.84 | 2.43 | 2.82 | 2.79 | 2.50 |
| $\gamma$ (mJ mol$^{-1}$ K$^{-2}$) | 12.12 | 13.94 | | 13.37 | 12.26 | 12.05 |
| $\beta$ (mJ mol$^{-1}$ K$^{-4}$) | 2.20 | 2.71 | | 1.96 | 1.87 | 1.97 |
| $\Theta_D$ (K) | 183.5(1) | 171.1(1) | | 190.6(2) | 193.6(2) | 190.3(1) |
| $\Delta C/\gamma T_c$ | 1.58 | 1.44 | | 1.45 | 1.51 | 1.50 |
| $\lambda_{ep}$ | 0.66 | 0.68 | | 0.66 | 0.65 | 0.63 |
| $N(E_F)$ (states/eV f.u) | 3.13 | 3.52 | | 3.41 | 3.15 | 3.10 |
| $-dH_{c2}/dT$ (T/K) | 0.105 | 0.107 | 0.098 | 0.100 | 0.125 | 0.066 |
| $\mu_0 H_{c2}$(T) | 0.202 | 0.212 | 0.167 | 0.198 | 0.247 | 0.12 |
| $\mu_0 H^P$(T) | 5.12 | 5.28 | 4.52 | 5.26 | 5.24 | 4.65 |
| $\mu_0 H_{c1}$(T) | 0.060 | 0.095 | 0.045 | 0.062 | 0.098 | 0.028 |
| $\xi_{GL}(0)$ (nm) | 40.4 | 39.3 | 44.4 | 40.7 | 36.3 | 52.8 |

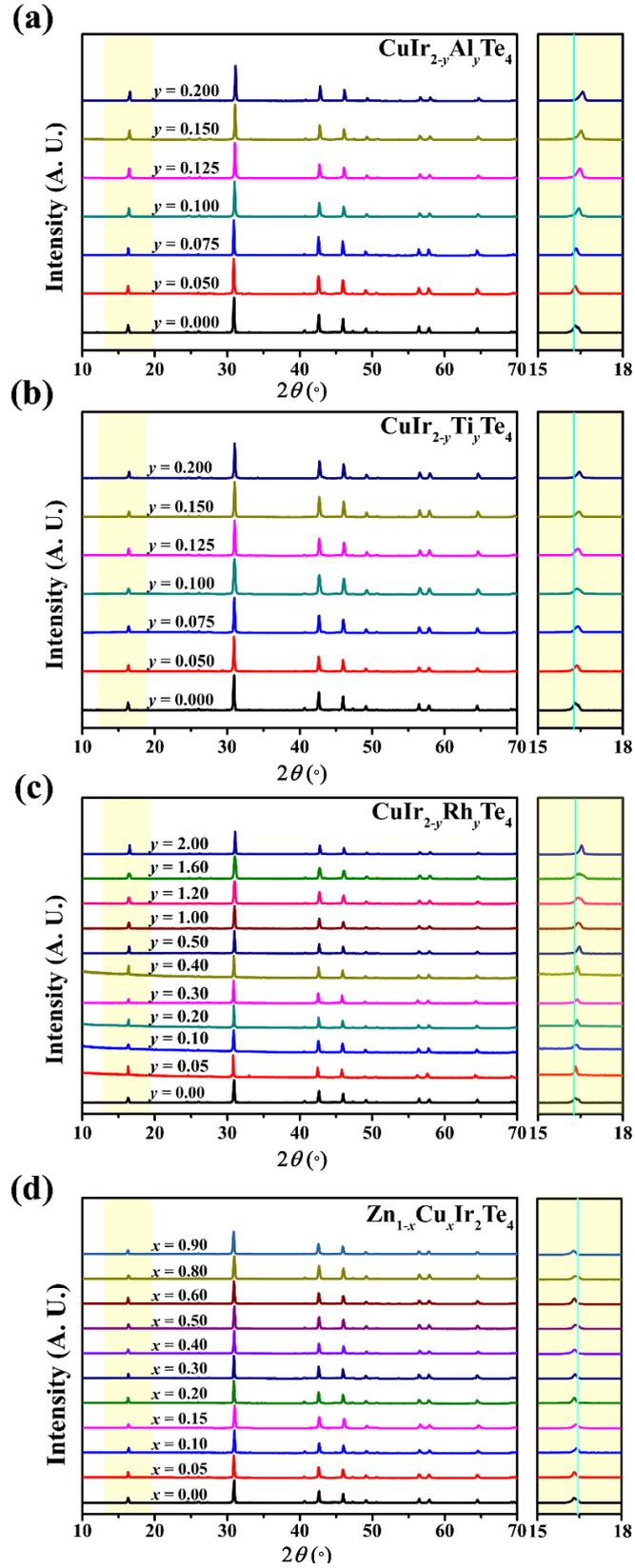

**Figure 1**. Powder XRD patterns (Cu Kα) for (a) CuIr$_{2-y}$Al$_y$Te$_4$ (0.0 ≤ $y$ ≤ 0.20), (b) CuIr$_{2-y}$Ti$_y$Te$_4$ (0.0 ≤ $y$ ≤ 0.20), (c) CuIr$_{2-y}$Rh$_y$Te$_4$ (0.0 ≤ $y$ ≤ 2.00) and (d) Zn$_x$Cu$_{1-x}$Ir$_2$Te$_4$ (0.0 ≤ $x$ ≤ 0.90). Inset shows the enlargement of peak (001).

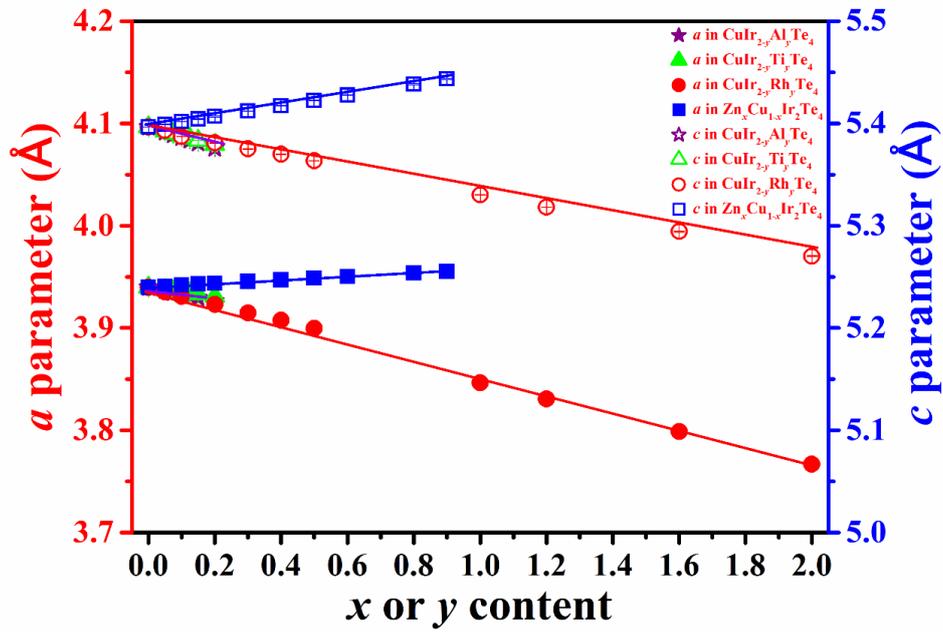

**Figure 2**. The evolution of lattice parameters $a$ and $c$ of CuIr$_{2-y}$Al$_y$Te$_4$ ($0.0 \leq y \leq 0.20$), CuIr$_{2-y}$Ti$_y$Te$_4$ ($0.0 \leq y \leq 0.20$), CuIr$_{2-y}$Rh$_y$Te$_4$ ($0.0 \leq y \leq 2.00$) and Zn$_x$Cu$_{1-x}$Ir$_2$Te$_4$ ($0.0 \leq x \leq 0.90$).

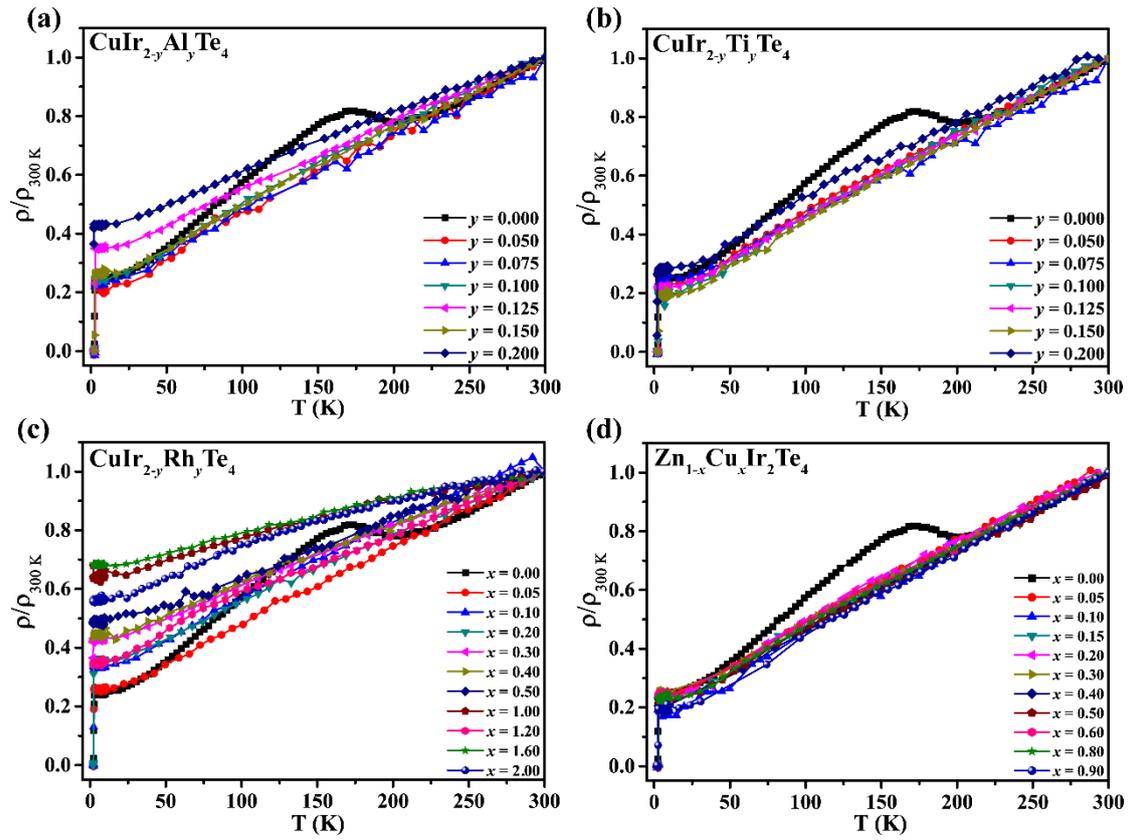

**Figure 3.** The temperature dependence of the resistivity ratio ($\rho/\rho_{300K}$) for the polycrystalline (a) $CuIr_{2-y}Al_yTe_4$ ($0.0 \leq y \leq 0.20$), (b) $CuIr_{2-y}Ti_yTe_4$ ($0.0 \leq y \leq 0.20$), (c) $CuIr_{2-y}Rh_yTe_4$ ($0.0 \leq y \leq 2.00$) and (d) $Zn_xCu_{1-x}Ir_2Te_4$ ($0.0 \leq x \leq 0.90$).

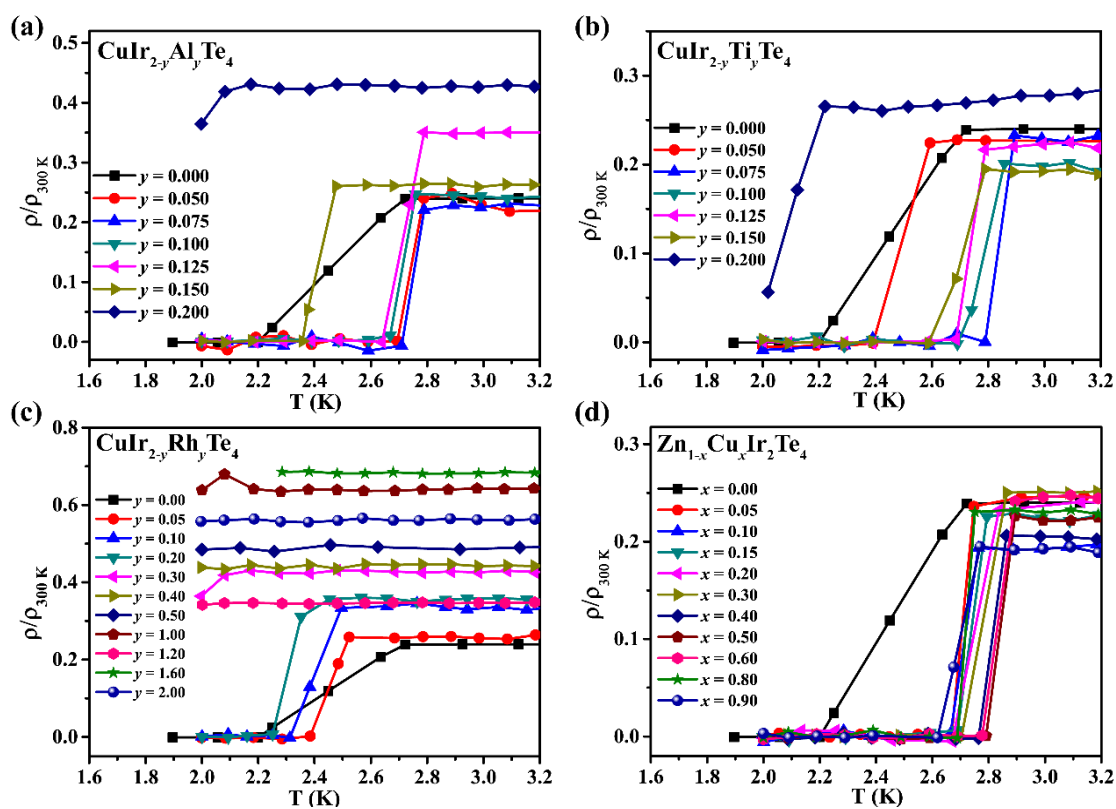

**Figure 4.** The temperature dependence of the resistivity ratio ($\rho/\rho_{300K}$) for the polycrystalline (a) CuIr$_{2-y}$Al$_y$Te$_4$ (0.0 ≤ $y$ ≤ 0.20), (b) CuIr$_{2-y}$Ti$_y$Te$_4$ (0.0 ≤ $y$ ≤ 0.20), (c) CuIr$_{2-y}$Rh$_y$Te$_4$ (0.0 ≤ $y$ ≤ 2.00) and (d) Zn$_x$Cu$_{1-x}$Ir$_2$Te$_4$ (0.0 ≤ $x$ ≤ 0.90) samples at low temperature from 1.6 to 3.2 K.

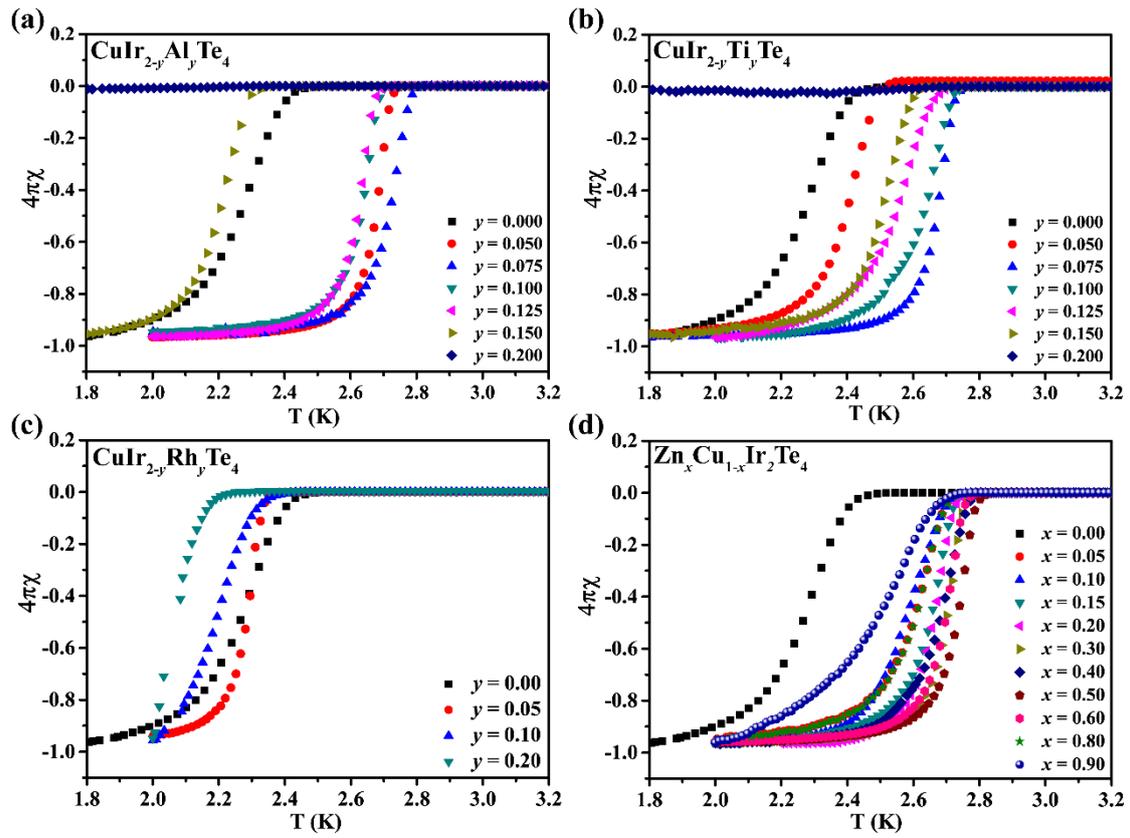

**Figure 5.** Magnetic susceptibilities for (a) CuIr$_{2-y}$Al$_y$Te$_4$ ($0.0 \leq y \leq 0.20$), (b) CuIr$_{2-y}$Ti$_y$Te$_4$ ($0.0 \leq y \leq 0.20$), (c) CuIr$_{2-y}$Rh$_y$Te$_4$ ($0.0 \leq y \leq 2.00$) (c) and (d) Zn$_x$Cu$_{1-x}$Ir$_2$Te$_4$ ($0.0 \leq x \leq 0.90$) at the superconducting transitions with applied DC field 20 Oe.

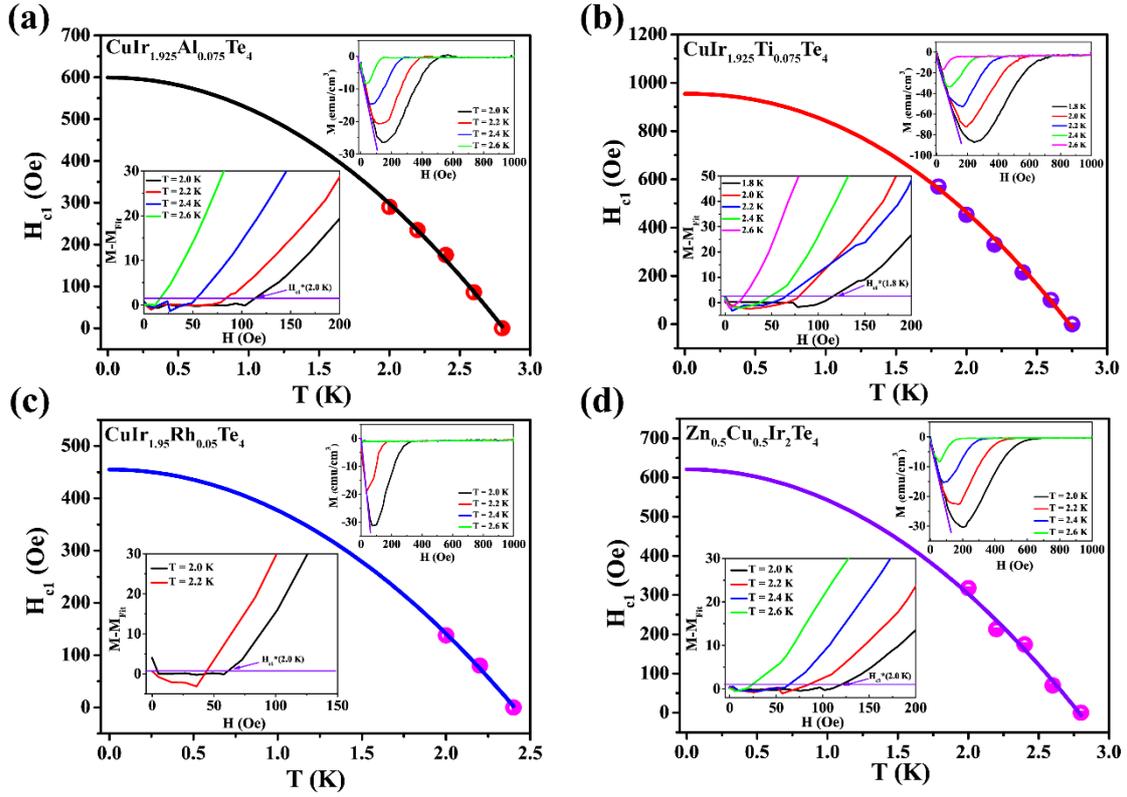

**Figure 6.** Temperature dependence of the lower critical fields ($\mu_0 H_{c1}$) for the representative (a) $CuIr_{1.925}Al_{0.075}Te_4$, (b) $CuIr_{1.925}Ti_{0.075}Te_4$, (c) $CuIr_{1.95}Rh_{0.05}Te_4$ and (d) $Zn_{0.5}Cu_{0.5}Ir_2Te_4$ samples. Up right corner inset shows magnetic susceptibility at low applied magnetics fields at various temperatures for each compound, bottom left inset shows M-M$_{fit}$ *vs* H.

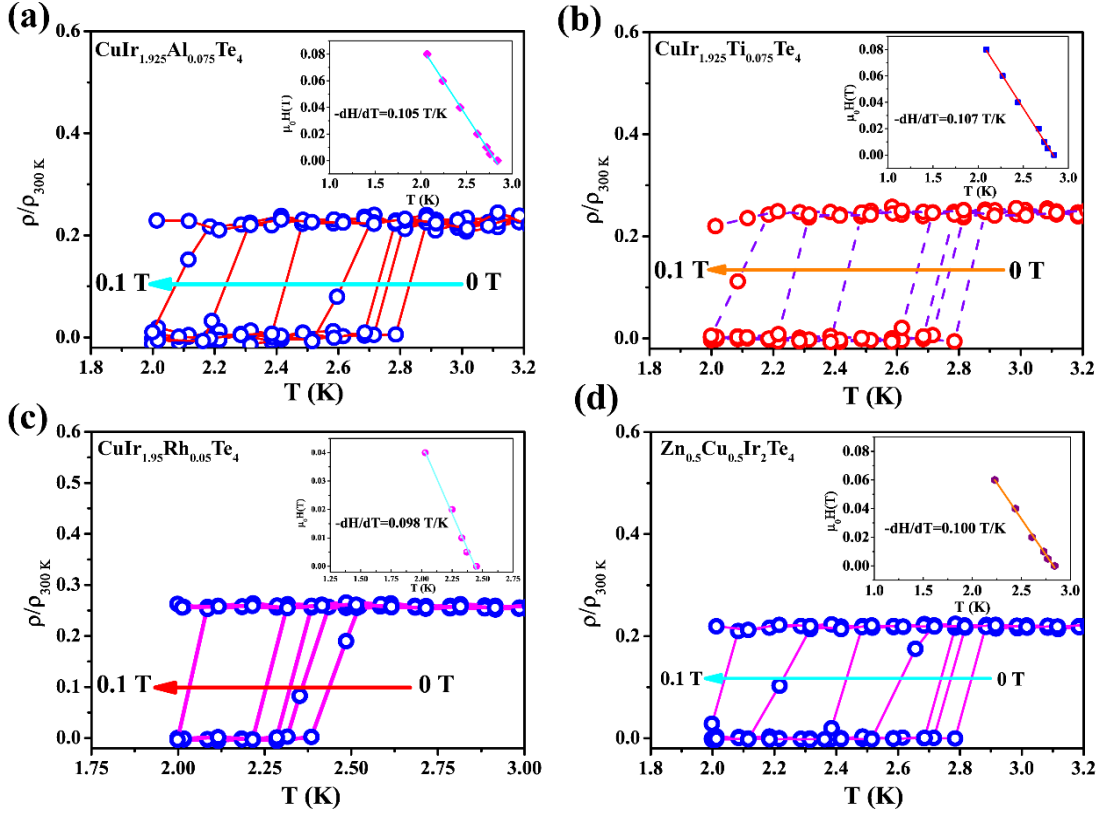

**Figure 7.** Low temperature resistivity at various applied fields for the representative (a) CuIr$_{1.925}$Al$_{0.075}$Te$_4$, (b) CuIr$_{1.925}$Ti$_{0.075}$Te$_4$, (c) CuIr$_{1.95}$Rh$_{0.05}$Te$_4$ and (d) Zn$_{0.5}$Cu$_{0.5}$Ir$_2$Te$_4$ samples. Inset shows $\mu_0H(T)$ at different $T_c$, red solid line shows linearly fitting to the data to estimate $\mu_0H_{c2}$.

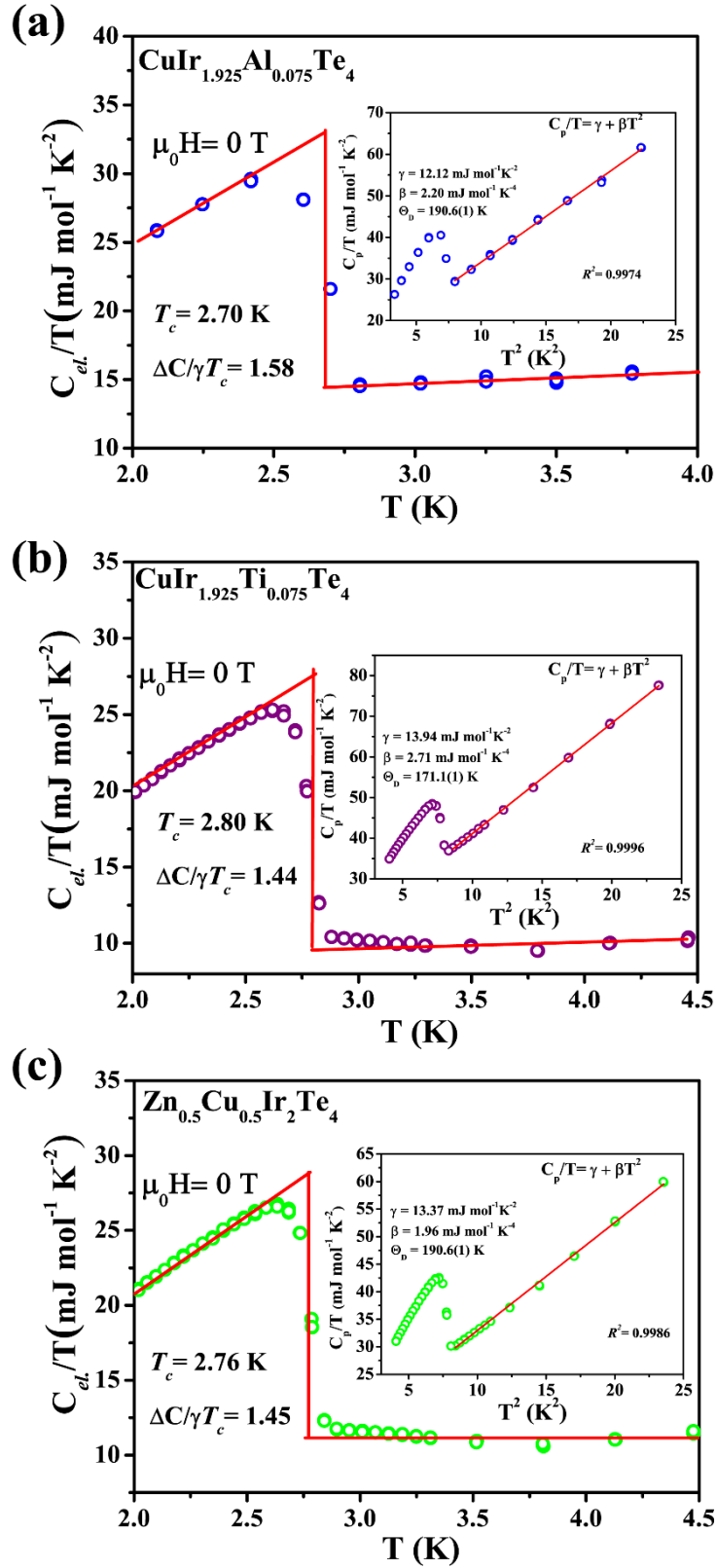

**Figure 8**. Electronic heat capacity divided by temperature ($C_{el.}$/T) *vs*. T measured at low temperature under zero applied magnetic field for the representative (a) $CuIr_{1.925}Al_{0.075}Te_4$, (b) $CuIr_{1.925}Ti_{0.075}Te_4$ (b) and (c) $Zn_{0.5}Cu_{0.5}Ir_2Te_4$. Inset: $C_p$/T *vs*. $T^2$ shown for the low temperature region and fitted to a line.

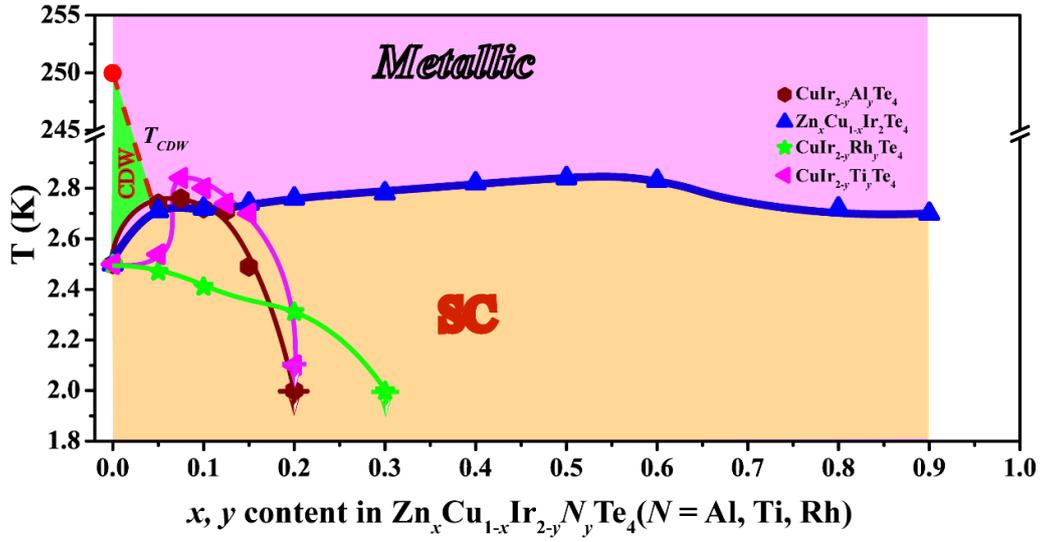

**Figure 9.** The electronic phase diagram for CuIr$_{2-y}$Al$_y$Te$_4$ (0.0 ≤ $y$ ≤ 0.20), CuIr$_{2-y}$Ti$_y$Te$_4$ (0.0 ≤ $y$ ≤ 0.20), CuIr$_{2-y}$Rh$_y$Te$_4$ (0.0 ≤ $y$ ≤ 2.00) and Zn$_x$Cu$_{1-x}$Ir$_2$Te$_4$ (0.0 ≤ $x$ ≤ 0.90). $T_c$ was all obtained via resistivity measurements.

# Supporting Information

# Charge density wave and superconductivity in the family of telluride chalcogenides $Zn_{1-x}Cu_xIr_{2-y}N(N = Al, Ti, Rh)_yTe_4$

Dong Yan[a], Yijie Zeng[b], Yishi Lin[c], Lingyong Zeng[a], Junjie Yin[b], Yuan He[a], Mebrouka Boubeche[a], Meng Wang[b], Yihua Wang[c], Dao-Xin Yao[b], Huixia Luo[a*]

[a]*School of Materials Science and Engineering and Key Lab Polymer Composite & Functional Materials, Sun Yat-Sen University, No. 135, Xingang Xi Road, Guangzhou, 510275, P. R. China*
[b]*School of Physics, State Key Laboratory of Optoelectronic Materials and Technologies, Sun Yat-Sen University, No. 135, Xingang Xi Road, Guangzhou, 510275, P. R. China*
[c]*Department of Physics, Fudan University, Shanghai, 200433, China*

[*]*Corresponding author/authors complete details (Telephone; E-mail:) (+0086)-2039386124*
*luohx7@mail.sysu.edu.cn*

---

**\*** [*]*Corresponding author/authors complete details (Telephone; E-mail:) (+0086)-2039386124, luohx7@mail.sysu.edu.cn*

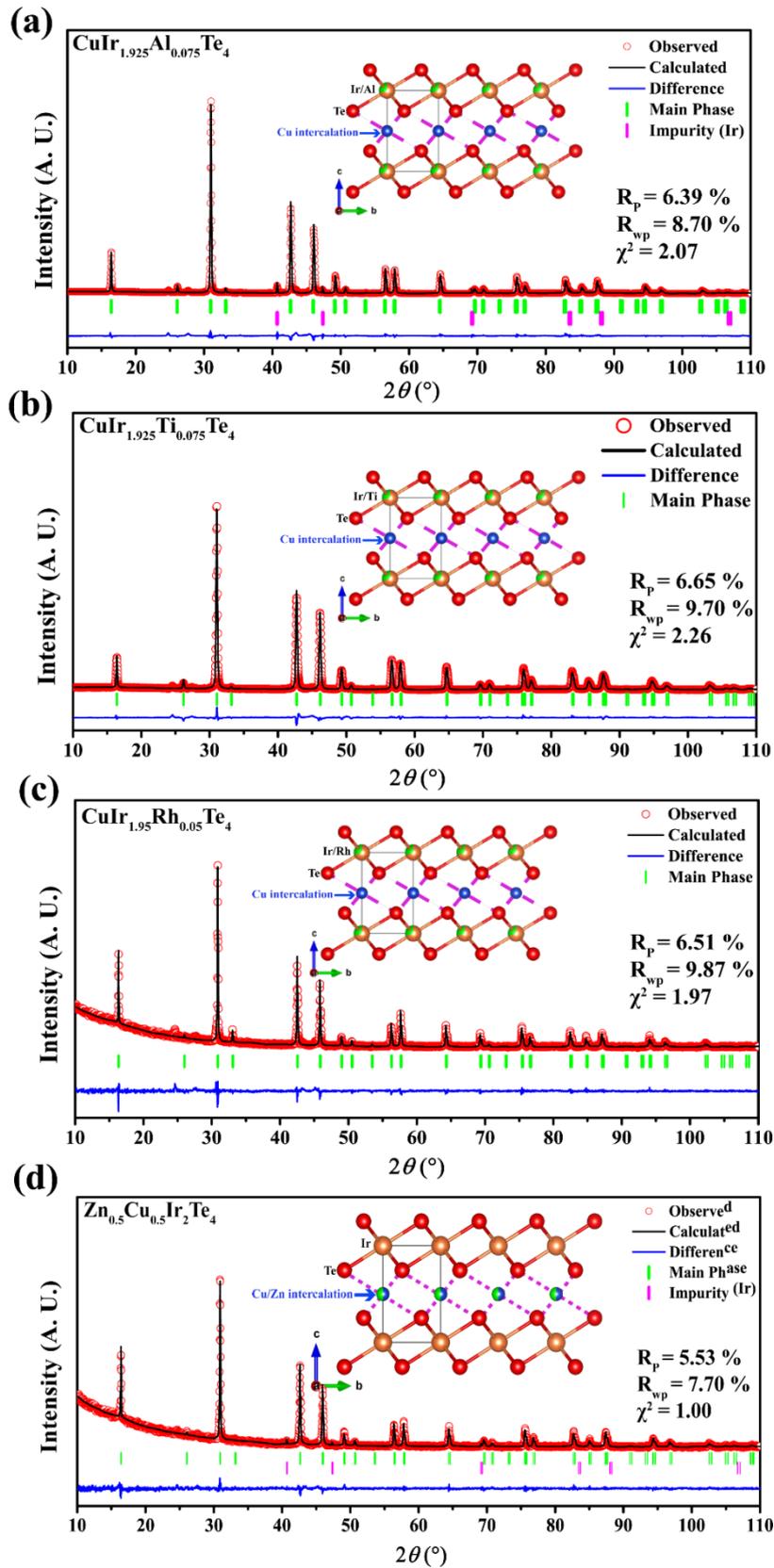

**Figure S1**. The refinements of representative $CuIr_{1.925-x}Al_{0.075}Te_4$, $CuIr_{1.925}Ti_{0.075}Te_4$, $CuIr_{1.95}Rh_{0.05}Te_4$ and $Zn_{0.5}Cu_{0.5}Ir_2Te_4$ polycrystalline samples.